\documentclass[amsmath,amssymb,amsbsy,reprint,prb,preprintnumbers,showpacs,superscriptaddress]{revtex4-2}
\usepackage{graphicx,color}
\usepackage{dcolumn}
\usepackage{bm}
\usepackage{braket}
\usepackage{mathtools}
\usepackage{ulem}
\usepackage[breaklinks,colorlinks=true,linkcolor=blue,urlcolor=blue,citecolor=blue]{hyperref}
\usepackage{times}
\usepackage{physics}
\usepackage{latexsym}
\usepackage{amsmath, amssymb}
\usepackage{mathtools}
\usepackage{multirow}

\newcommand{\be}{\begin{eqnarray}}
\newcommand{\ee}{\end{eqnarray}}

\begin{document}

\title{Strong-to-weak symmetry breaking states in stochastic dephasing stabilizer circuits}
\date{\today}
\author{Yoshihito Kuno} 
\thanks{These authors equally contributed}
\affiliation{Graduate School of Engineering Science, Akita University, Akita 010-8502, Japan}
\author{Takahiro Orito}
\thanks{These authors equally contributed}
\affiliation{Institute for Solid State Physics, The University of Tokyo, Kashiwa, Chiba, 277-8581, Japan}
\author{Ikuo Ichinose} 
\thanks{A professor emeritus}
\affiliation{Department of Applied Physics, Nagoya Institute of Technology, Nagoya, 466-8555, Japan}


\begin{abstract} 
Discovering mixed state quantum orders is an on-going issue.   
Recently, it has been recognized that there are (at least) two kinds of symmetries in the mixed state; strong and weak symmetries. Under symmetry-respective decoherence, 
spontaneous strong-to-weak symmetry breaking (SSSB) can occur. 
This work provides a scheme to describe SSSB and other decoherence phenomena in the mixed state by employing the stabilizer formalism and the efficient numerical algorithm of Clifford circuits.  
We present two systematic numerical studies. 
In a two-dimensional (2D) circuit with a stochastic Ising type decoherence, an SSSB phase transition is clearly observed and its criticality is elucidated by the numerical methods. 
In particular, we calculate R\'{e}nyi-2 correlations and estimate critical exponents of the SSSB transition.
For the second system, we introduce an idea of subgroup SSSB. 
As an example, we study a system with symmetry-protected-topological (SPT) order provided by both one-form and zero-form symmetries, and observe how the system evolves under decoherence. 
After displaying numerical results, we show that viewpoint of percolation is quite useful to understand the SSSB transition, which is applicable for a wide range of decohered states.
Finally, we comment on SSSB of one-form-symmetry exemplifying toric code.
\end{abstract}


\maketitle
\section{Introduction} 
Quantum decoherence is inevitable in realistic quantum systems \cite{Gardiner}, which
is induced by couplings between environment and quantum states. 
As a result, the states lose pure-state coherence and reduce to mixed states.  
Such decoherence may be regarded as an undesired effect in many quantum systems. 
Certainly, quantum devices such as quantum computer and quantum memory are significantly affected by quantum noise \cite{Preskill,Dennis2002,Wang2003}.  
However, such an effect can enhance the versatility of the quantum many-body state, that is, decoherence effects lead to various non-trivial quantum many-body states, properties of which are beyond the pure state picture in isolated quantum systems and the ordinary thermal mixed states.

As an example of the above observation, quantum measurements on quantum systems can induce various fruitful phenomena and drastic changes in the quantum state. 
By applying suitably chosen quantum measurements on a target quantum system, the backaction of quantum measurements renders the state exotic quantum matter, such as a long-range entanglement state \cite{Raussendorf2001,Raussendorf2005,Verresen2021_cat,Tantivasadakarn2022,Lee2022,Lu2023,Chen2023_v1,Zhu2023,Angelidi_2023,KOI2024_2}, topological states of matter \cite{Lavasani2021,Lavasani2021_2,Klocke2022,Negari2023,KI2023,Morral-Yepes2023,KOI2024,OKI2024} and highly entanglement states \cite{Ippoliti2021,Sriram2023,Lavasani2023,KOI2023}.  
In this sense, quantum measurement can be regarded as an important tool to manipulate quantum many-body systems, especially for measurement-based quantum computation (MBQC) \cite{Raussendorf2001,Raussendorf2006,Briegel2009,Wei_rev2018}. 

Identifying and characterizing various orders in mixed quantum many-body states is now an on-going issue in condensed matter physics and quantum information theory.  
As one of the attracting phenomena, topological orders survive even in systems at finite temperature and/or under decoherence  \cite{T_C_Lu2020,Y_H_Chen2024} or they tend to acquire another symmetry without losing topological properties \cite{Zhang2022}. 
Similarly, some of the states with symmetry-protected-topological order (SPT) \cite{Chen_SPT,Pollmann2012} survive under decoherence \cite{J_Y_Lee2022,Zhang2024}, and sometimes they change to average SPT state as studied in Ref.~\cite{Ma2023}. 

Categorizing mixed states emerging due to decoherence has attracted a lot of attention in recent years. 
In general, manipulating mixed states is not directly an easy task, and previous studies utilize some kinds of mapping from mixed states to pure states. 
Some of them are the doubled Hilbert state formalism and purification \cite{Choi_map1,Choi_map2} as well as effective field theory methods from a coarse-grained point of view \cite{Tantivasadakarn2021,J_Y_Lee2022}. 
In this study, we propose and examine another prescription investigating mixed states under decoherence, that is, projective stabilizer circuits. 
By using the circuit under decoherence, mixed states can be treated faithfully and physical quantities such as correlation functions can be calculated numerically even if a target system is not tractable by analytical methods and cannot be mapped to well-known models.
In some cases including systems studied in this work, a qualitative picture from the geometrical viewpoint can be readily obtained, which helps us to understand numerical results. 
Symmetry aspect of decohered mixed states is under intensive study, in particular, strong and weak symmetries and a spontaneous breaking transition between them.
In this paper, we shall apply the projective stabilizer circuit to that issue to clarify how systems undergo a certain symmetry phase transition and investigate the critical behavior.

Symmetry-breaking phenomena of the mixed state are highly attractive.
The notion of spontaneous symmetry breaking in pure states can be extended to quantum mixed states.  
However, for a general mixed state represented by a density matrix, two types of symmetry, i.e., strong and weak symmetries, can be considered and discriminated \cite{Groot}. 
Very recently, the possibility of spontaneous strong-to-weak symmetry breaking (SSSB) was proposed and studied \cite{J_Y_Lee2023,Lessa2024,Sala2024}. 
That is, the strong symmetry of mixed states can be spontaneously broken to weak symmetry (or no symmetry) by decoherence through channels with strong symmetry.
It is expected that this phenomenon emerges in various open quantum systems. 
Although the concept of the SSSB is very interesting and it is characterized by some physical quantities
in specific models \cite{Lessa2024,Sala2024}, a more general formulation providing numerical methods and examples is desired. 
As we mentioned in the above, the projective stabilizer circuit can be such a formalism.
This formalism can be used for solving the following issues; what kind of quantum channels
(decoherence and dephasing) cause SSSB, and for a channel of decoherence in which
 an SSSB emerges, whether a phase transition takes place there
by controlling the strength of decoherence and if so, how the criticality of that SSSB transition looks like. 

This work gives concrete examples of the projective stabilizer circuit including the numerical verification for the analytical observations and detailed investigation of the SSSB. 
We first explain the concept of the SSSB by following recent works and then show various concrete examples of SSSB states and corresponding strongly-symmetry channels in our formalism including subgroup SSSB states.
There, all of the channels are composed of projective measurements (maximal decoherence limit), that is, outcomes of measurements are discarded and not recorded. 
In the main part of the study, we consider a circuit dynamics with local stochastic dephasing called {\it dephasing-only} circuit (DoC), which relies on a quantum trajectory picture of mixed states. 
We numerically demonstrate the SSSB in the averaged picture over density matrix trajectories, where we employ large-scale stabilizer simulation \cite{Gottesman1997,Aaronson2004}. 

The rest of this paper is organized as follows. 
In Sec.~II, we introduce the concept of the SSSB by following the previous works and its characterization.  
In Sec.~III, we show some concrete examples of `perfect' SSSB created by the single-layer dephasing channel. 
In Sec.~IV, we move to study on stochastic DoC systems by using the efficient stabilizer algorithm for large system sizes. 
We present an SSSB phase transition in two kinds of two-dimensional (2D) systems. 
The first one is defined on a 2D square lattice system and the second is on a Lieb lattice. 
We verify the phase transition induced solely by local dephasing and investigate its criticality by using numerical large-scale calculations.  
The notion of a cluster stabilizer plays an important role there.
In Sec.~VI, we study the relation between the stabilizer formalism of the DoC and percolation to obtain
a qualitative understanding of the numerically obtained results.
Section VII is devoted to conclusion and discussion.

\section{Basics of SSSB}
We start to briefly explain the notion of the SSSB by following Refs.~\cite{J_Y_Lee2023,Lessa2024,Sala2024}. 
The SSSB is a novel type of quantum order defined on mixed states, which is an extended notion of spontaneous symmetry breaking (SSB) \cite{Altland_Simons} in pure states. 

Let us first introduce the strong and weak symmetries for a density matrix.
In general, a density matrix (mixed state) can have two types of symmetries in principle. 
One is the strong symmetry \cite{Groot}
\begin{eqnarray}
U_g\rho=e^{i\theta}\rho,
\end{eqnarray}
where $\rho$ is a mixed state and $U_g$ is a symmetry operation of an element $g$ of a symmetry group $G$ and $\theta$ is a certain global phase factor. 

Next, the other is weak symmetry, which is an ordinary one  and defined as  
\begin{eqnarray}
U_{g}\rho U^\dagger_{g} =\rho.
\end{eqnarray}
This condition is called the average or weak symmetry 
condition \cite{Ma2023}, where the symmetry is satisfied after taking the ensemble average in general.

Strong and weak symmetry conditions are also defined for quantum channels. 
Generic quantum channel is given by completely positive trace preserving (CPTP) maps. 
The operator-sum representation of the channel is given as
\cite{Nielsen_Chuang}
\begin{eqnarray}
\mathcal{E}(\rho)=\sum^{N-1}_{\ell=0}K_{\ell} \rho K^\dagger_{\ell},
\end{eqnarray}
where $K_{\ell}$ is a Kraus operator satisfying $\sum^{N-1}_{\ell=0} K^\dagger_{\ell} K_{\ell}=I$. 
The quantum channel $\mathcal{E}$ induces changes in general mixed states including non-unitary transformations such as decoherence and quantum measurements.

Here, the strong symmetry condition on the channel for a symmetry $G$ is represented as \cite{Groot}
$$K_{\ell}U_g=e^{i\theta} U_g K_{\ell}$$
for any $\ell$ and $g \in G$, where $\theta$ is a single phase.
On the other hand, weak symmetry condition on the channel for a symmetry $G$ is  given as 
\begin{eqnarray}
U_g\biggl[\sum_{\ell}K_{\ell} \rho K^\dagger_{\ell}\biggr]U^\dagger_g=\mathcal{E}(\rho).
\end{eqnarray}
That is, each Kraus operator itself is not commutative with $U_g$. 

Based on the notion of the strong symmetry, let us explain the SSSB. 
Here, we limit the discussion within a unitary on-site strong symmetry, the operation of which is represented by $U=\otimes_j U_j$, where $U_j$ is an on-site charge operator. 
We consider a mixed state $\rho$ with spontaneously-broken strong symmetry.
Then we consider the conventional correlation function of the charged operator for $U_j$, 
\begin{eqnarray}
C^{\rm I}_{O_i O^\dagger_j}(\rho)\equiv{\rm Tr}[\rho O_i O^\dagger_j], \:[\forall i,j\ (i\neq j)]\nonumber
\end{eqnarray}
where $O_i$ is a local charged operator for the symmetry (when $U_j$ and $O_j$ are elements in Pauli group, $\{O_j,U_j\}=0$). 
Then, the above has exponential decay for the distance $|i-j|$ and no long-range orders \cite{Lessa2024}, 
\begin{eqnarray}
\lim_{|i-j|\to \infty}C^{\rm I}_{O_i O^\dagger_j}(\rho)=0.
\label{cond1}
\end{eqnarray}

This means that the conventional SSB does \textit{not} occur in an SSSB state.

On the other hand in some cases, for a mixed state with strong symmetry 
we consider the R\'{e}nyi-2 correlator defined by~\cite{Lessa2024},
\begin{eqnarray}
C^{\rm II}_{O_i O^\dagger_j}(\rho)\equiv \frac{{\rm Tr}[O_i O^\dagger_j\rho O_i O^\dagger_j \rho ]}{{\rm Tr}[\rho^2]}, \:[\forall i,j\ (i\neq j)].\nonumber
\end{eqnarray}
The R\'{e}nyi-2 correlator gives one of the criteria for the emergence of SSSB \cite{Lessa2024}. The state with SSSB has a non-vanishing long-range order,
\begin{eqnarray}
\lim_{|i-j|\to \infty} C^{\rm II}_{O_j O^\dagger_i}(\rho)=\mathcal{O}(1).
\label{cond2}
\end{eqnarray}
That is, when a mixed state $\rho$ with the strong symmetry satisfies the conditions of Eqs.~(\ref{cond1}) and (\ref{cond2}), the state $\rho$ is an SSSB state. 
The simplest SSSB state is a glassy GHZ mixed state $\rho_{\rm GHZ}=\frac{I+\prod_j X_j}{2}$ \cite{stab_GHZ}, where the symmetry is the parity $P=\prod X_j$.
One can easily check that the state $\rho_{\rm GHZ}$ satisfies the conditions of Eqs.~(\ref{cond1}) and (\ref{cond2}).

Here, we should note another diagnosis to identify the SSSB dubbed fidelity correlator \cite{Lessa2024}. The SSSB state characterized by the fidelity correlator has stability theorem, indicating possiblity to identify more broader SSSB states than those by the R\'{e}nyi-2 correlator \cite{Lessa2024}. However, the R\'{e}nyi-2 correlator acts as a sufficient indicator to the SSSB state, and the calculation of the R\'{e}nyi-2 correlator is numerically more tractable than that of the fidelity correlator. Thus, we focus on R\'{e}nyi-2 correlator in this work.

Based on the definition of SSSB state, 
we are interested in how and which quantum channel changes an initial strongly symmetric state without SSSB into an SSSB mixed state.
To this end, the channel must be a strongly-symmetric one. 
That is, under the following state-changing operation
$\mathcal{E}$,   
\begin{eqnarray}
&&\rho_{\rm ini} \xlongrightarrow{\mathcal{E}} \rho_{\rm fin},\\
&&C^{\rm II}_{O_i O^\dagger_j}(\rho_{\rm ini})=0,\;\; 
C^{\rm II}_{O_i O^\dagger_j}(\rho_{\rm fin})=\mathcal{O}(1), \nonumber
\end{eqnarray}
where ${\rm Tr}[\rho_{{\rm ini}({\rm fin})} O_i O^\dagger_j]=0$.
We call such a channel $\mathcal{E}$ an SSSB-inducing channel.
To obtain concrete examples of the SSSB-inducing channel is one of main objects in this work.
Furthermore, after showing examples of such a channel, we address the problem if a kind of phase transition takes place under controlling the strength of decoherence of the channel.
In our formalism, the efficient numerical study is available.

\section{Concrete example of SSSB state for maximum dephasing limit}
In this section, we shall show some typical examples of the SSSB-inducing channel. 
Here, by making use of the stabilizer formalism \cite{Nielsen_Chuang} and dephasing channel, 
we elucidate the change of the initial states to SSSB states through a sequence of maximal dephasing,
which is explained shortly. 
The obtained observation gives insight into the study on the SSSB phase transition in stochastic DoCs, which is studied in Sec.~IV. 
[The basic standard rules of stabilizer generators and corresponding state changes in channels are explained in Appendices A and B.]

\subsection{$Z_2$-Ising SSSB}
We first discuss the $Z_2$-Ising SSSB state and its corresponding 
SSSB-inducing channels in 1D and 2D systems.\\

\noindent \underline{1D case}: 
Let us consider 1D $L$ qubit chain with periodic boundary conditions and set the initial density matrix as $+X$ product pure state, 
 $\rho_{\rm ini}\longleftrightarrow S_{\rm ini}=\{X_\ell | \ell=0,L-1\}$, where $S_{\rm ini}$ is the corresponding set of stabilizer generators to $\rho_{\rm ini}$. 
 The target symmetry is $Z_2$ parity $G=\prod^{L-1}_{j=0}X_j$. 
Then, consider a single-round link $Z_\ell Z_{\ell+1}$ dephasing channel, which is strong symmetric to $G$:
\begin{eqnarray}
&&\mathcal{E}^{ZZ_{NN}}_{\rm all}[\rho]= \prod^{L-1}_{i=0} \mathcal{E}^{Z_iZ_{i+1}}[\rho],\\
&&\mathcal{E}^{Z_i Z_{i+1}}[\rho]=\sum_{\beta_i=\pm}P^{Z_i Z_{i+1}}_{\beta_i}\rho P^{Z_iZ_{i+1}}_{\beta_i}.
\label{entire_dephasing_channel_ZZ}
\end{eqnarray}
This channel represents a projective measurement without monitoring (recording) outcomes and $P^{Z_iZ_{i+1}}_{\beta_i}=\frac{1+\beta_i Z_iZ_{i+1}}{2}$. 
Then, the finial mixed state denoted by $\rho_{\rm fin}$ is given by 
$\mathcal{E}^{ZZ_{NN}}_{\rm all}[\rho_{\rm ini}]=\frac{I+\prod_j X_j}{2}$. 
This is easily obtained by the dephasing update procedure of the set of the stabilizer generators with the efficient stabilizer algorithm shown in Appendix B.
Explicitly, we show the dephasing process:
\be
&&\mathcal{E}^{Z_0Z_1}: \{X_0,X_1,X_2,X_3,\cdots\} \to \{X_0X_1,X_2,X_3,\cdots\}  \nonumber \\
&&\mathcal{E}^{Z_1Z_2}: \{X_0X_1,X_2,X_3,\cdots\} \to \{X_0X_1X_2,X_3,\cdots\},  \nonumber
\ee
etc.
It is interesting to observe that the $(Z_iZ_{i+1})$-dephasing operation works as an eliminator for an anti-commute generator inducing mixing of the state and is regarded as glue which merges $(X_i,X_j)$ in the stabilizer group.
Then, we obtain straightforwardly
$S_{\rm ini}\xlongrightarrow{\mathcal{E}^{Z_0Z_1}} \cdots \xlongrightarrow{\mathcal{E}^{Z_{L}Z_{0}}} S_{\rm fin}=\{\prod^{L-1}_{\ell=0}X_{\ell}\}$ \cite{stab_1DIsing}. 
The final mixed state corresponding to $S_{\rm fin}$ is nothing but the glassy GHZ state. 
Note here that the final mixed state has only one stabilizer generator corresponding to the generator of the target symmetry generator. 
Then, we call the above prescription $\mathcal{E}_{\rm all}$ maximal dephasing (decoherence) limit.
The presence of the single stabilizer generator gives a finite value of R\'{e}nyi-2 correlator $C^{\rm II}_{Z_i Z_j}(\rho_{\rm fin})=1$ while $C^{\rm II}_{Z_i Z_j}(\rho_{\rm ini})=0$ with 
$C^{\rm I}_{Z_i Z_j}(\rho_{\rm ini(fin)})=0$. 
Thus, the state $\rho_{\rm fin}$ is SSSB. \\

\noindent \underline{2D case}: By using the same way as that of the 1D case, a 2D system also exhibits $Z_2$-Ising SSSB. 
Let us consider 2D $L_x\times L_y$ square lattice with periodic boundary conditions and set the initial density matrix as $+X$ product pure state, 
$\rho_{\rm ini}\longleftrightarrow S_{\rm ini}=\{X_{(i_x,i_y)} | i_{x(y)}=0,1,\cdots, L_{x(y)}-1 \}$. 
Again, the target symmetry is $Z_2$ parity $G=\prod_{all: (i_x,i_y)}X_{(i_x,i_y)}$. 
Then, consider a single-round link $ZZ$-dephasing channel, which is strong symmetric to $G$,
\begin{eqnarray}
&&\mathcal{E}^{ZZ_{NN}}_{2D,{\rm all}}[\rho]= \prod_{(i_x,i_y)} \mathcal{E}^{Z_{(i_x,i_y)}Z_{(i_x+1,i_y)}}\circ\mathcal{E}^{Z_{(i_x,i_y)}Z_{(i_x,i_y+1)}}[\rho].\nonumber
\label{2D_entire_dephasing_channel_ZZ}
\end{eqnarray}
The schematic image of the local link ZZ-dephasing is shown in Fig.~\ref{Fig_lat} (a). 
Then, the final mixed state denoted by $\rho_{\rm fin}$ is given by 
$\mathcal{E}^{ZZ_{NN}}_{2D,{\rm all}}[\rho_{ini}]=\frac{I+\prod_{(i_x,i_y)} X_{(i_x,i_y)}}{2}$. 
This state is easily obtained 
$S_{\rm ini}\xlongrightarrow{\mathcal{E}^{Z_{(0,0)}Z_{(1,0)}}} \cdots \xlongrightarrow{\mathcal{E}^{Z_{(L_{x}-1,L_{y}-1)}Z_{(L_{x}-1,0)}}} S_{\rm fin}
=\{\prod_{{\rm all}: (i_x,i_y)}X_{(i_x,i_y)}\}$. 
The corresponding mixed state to $S_{\rm fin}$ is the 2D glassy GHZ state. 
This presence of the single stabilizer generator gives a finite value of R\'{e}nyi-2 correlator $C^{\rm II}_{Z_{(i_x,i_y)} Z_{(j_x,j_y)}}(\rho_{\rm fin})=1$ while $C^{\rm II}_{Z_{(i_x,i_y)} Z_{(j_x,j_y)}}(\rho_{\rm ini})=0$ with 
$C^{\rm I}_{Z_{(i_x,i_y)} Z_{(j_x,j_y)}}(\rho_{\rm ini(fin)})=0$. 
Thus, the state $\rho_{\rm fin}$ is SSSB. 
Here, note that from the above examples manipulated in the stabilizer formalism, the maximal SSSB states are mixed states stabilized only by symmetry generators.

\subsection{Subgroup SSSB}
When the target system has multiple strong symmetries, we can consider a partial breakdown of the symmetries, that is, some subgroup of the symmetries spontaneously breaks down to weak symmetry.
We call the above situation subgroup SSSB. 
Let us study two cases of subgroup SSSB as concrete examples.\\

\noindent \underline{Subgroup Ising SSSB on the 1D cluster state}:  
Let us consider the 1D cluster state defined on the periodic chain, $\rho_{\rm CS}=|\Psi_{\rm CS}\rangle\langle \Psi_{\rm CS}|$ as a pure initial state.  
The state $|\Psi_{\rm CS}\rangle$ is stabilized as $g^{CS}_{\ell}|\Psi_{\rm CS}\rangle=|\Psi_{\rm CS}\rangle$ where $g^{CS}_{\ell}$ is a $ZXZ$ stabilizer generator given by $g^{CS}_{\ell}=Z_{\ell}X_{\ell+1}Z_{\ell+2}$ ($\ell=0, 1,\cdots, L-1$).
A set of stabilizer generators are given by $\mathcal{S}^{CS}= \{g^{CS}_{0}, \cdots g^{CS}_{L-1}\}$. 
The 1D cluster state is a SPT state protected by $Z_2\times Z_2$ symmetry generated by $G_1=\prod^{L/2-1}_{j=0} X_{2j}$ and $G_2=\prod^{L/2-1}_{j=0} X_{2j+1}$.

Here, we consider the following $X$-dephasing channel. 
The local dephasing is given by
\begin{eqnarray}
\mathcal{E}^{X_i}[\rho]&=&\sum_{\beta_i=\pm}P^{X_i}_{\beta_i}\rho P^{X_i\dagger}_{\beta_i},
\label{local_X_dephasing_channel_0}
\end{eqnarray}
where $P^{X_i}_{\beta_i}=\frac{1+\beta_i X_i}{2}$. 
The channel $\mathcal{E}^{X_i}$ is strong symmetric to the $Z_2\times Z_2$ symmetry as it is given solely by $\{X_i\}$. Because the stabilizer generators preserving the short-range entanglement of the initial state are connected by the channel, a long-range correlation characterized by the R\'enyi-2 correlator is generated in the final mixed state.
Now we observe how the state $\rho_{\rm CS}$ evolves through the channel composed of  $\{\mathcal{E}^{X_i}\}$.\\

\noindent{\underline{(Case I) Entire system dephasing:}} 
We consider a single-layer entire-system-decoherence channel that is given by $
\mathcal{E}^{X}_{\rm all}[\rho]= \prod^{L-1}_{i=0} \mathcal{E}^{X}_i[\rho]$. 
Then, the final mixed state is given by
$\mathcal{S}^{CS}\xlongrightarrow{\mathcal{E}^{X}_{all}}
\{G_1,G_2\}
\Longrightarrow \rho^f_{\rm CS} = \frac{1}{2^L}(I+G_1+G_2+G_1G_2)$.
The decohered final state $\rho^f_{\rm CS}$ is stabilized solely by the symmetry generators of the initial SPT state \cite{inf_T}. 
As a result, the presence of the two stabilizer generators $\{G_1,G_2\}$ gives a finite value of R\'{e}nyi-2 correlator $C^{\rm II}_{Z_i Z_j}(\rho^f_{\rm CS})=\mathcal{O}(1)$ while $C^{\rm II}_{Z_i Z_j}(\rho_{\rm ini})=0$ with $C^{\rm I}_{Z_i Z_j}(\rho_{\rm ini(fin)})=0$. 
Thus, the state $\rho_{\rm fin}$ is SSSB. 
Note that the operator $Z_i Z_j$ in the R\'{e}nyi-2 correlator is a charged operator for the on-site $Z_2\times Z_2$ symmetry. 
The strong symmetric channel $\mathcal{E}^{X}_{\rm all}$ induces the SSSB state for the subgroup symmetries $G_1$ and $G_2$, which are the protection symmetries for the initial SPT state.\\

\noindent{\underline{(Case II) Even-site local dephasing:}} As a next example, consider even-sites dephasing given by $
\mathcal{E}^{X}_{\rm even}[\rho]= \prod^{L/2-1}_{j=0} \mathcal{E}^{X}_{2j}[\rho]$.
Let us apply $\{\mathcal{E}^{X}_{2j}\}$ for all even sites, then we obtain an updated set of stabilizer generators as follows, 
\begin{eqnarray}
&&\mathcal{S}^{CS}\xlongrightarrow{\mathcal{E}^{X}_{even}}\{G_2\}+\{ g^{CS}_{2\ell+1} |\ell=0,\cdots,L/2-1\} \nonumber\\
&&\Longrightarrow \rho^{f_e}_{CS}\equiv \frac{1}{2^{L/2-1}}\biggr[\frac{I+G_2}{2}\biggl]
\prod^{L/2-1}_{\ell=0}\biggr[\frac{1+g^{CS}_{2\ell+1}}{2}\biggr].
\end{eqnarray}
The decohered final state $\rho^{f_e}_{\rm CS}$ is stabilized by one of the on-site symmetry operators of the initial SPT state in addition to odd-site cluster $ZXZ$ generators. 
As a result, the presence of the single stabilizer generator $G_2$ gives a finite value of odd-site R\'{e}nyi-2 correlator $C^{\rm II}_{Z_{2i+1} Z_{2j+1}}(\rho^{f_e}_{\rm CS})=\mathcal{O}(1)$ while $C^{\rm II}_{Z_{2i} Z_{2j}}(\rho_{\rm CS})=0$ with 
$C^{\rm I}_{Z_{2i} Z_{2j}}(\rho_{\rm CS})=C^{\rm I}_{Z_{2i} Z_{2j}}(\rho^{f_e}_{\rm CS})=0$. 
This result indicates that the SSSB occurs on the odd-site subsystem. 
On the other hand, the even-site subsystem exhibits no SSSB as $C^{\rm II}_{Z_{2i} Z_{2j}}(\rho^{f_e}_{\rm CS})=0$. 

We get the following observation from the results in this subsection: When the SSSB or subgroup SSSB state appears and the generator of its target on-site symmetry is described by an element of Pauli group, then 
the set of the stabilizer generators always contains the generator of the target symmetry. 
This stabilizer induces non-zero R\'{e}nyi-2 correlator of a charged operator for the on-site symmetry.\\

\begin{figure}[t]
\begin{center} 
\vspace{0.5cm}
\includegraphics[width=8.8cm]{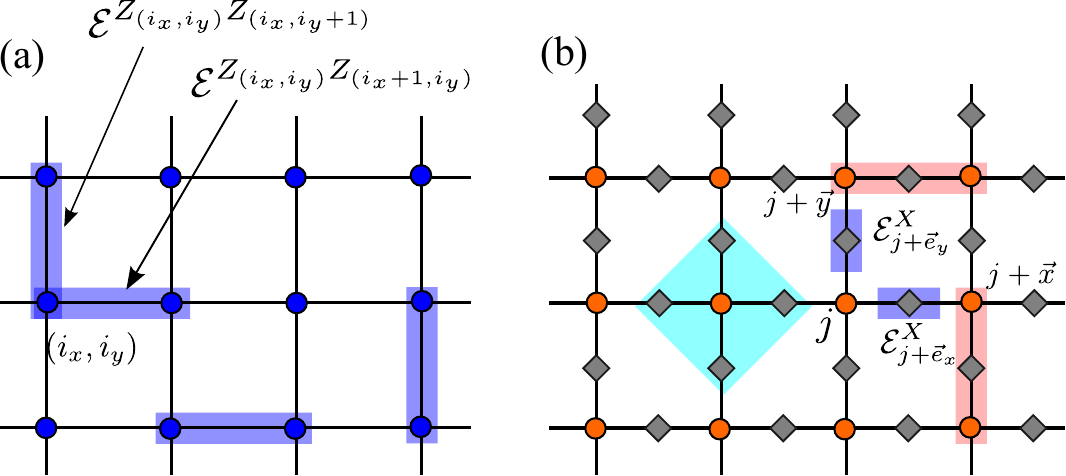}  
\end{center} 
\caption{(a) 2D square lattice. The blue shade bonds represent a link ZZ-dephasing with probability $r$.  
(b) Lieb lattice: The right blue diamond represents a vertex stabilizer $\tau^xZZZZ$ and the red shade represents the link stabilizer $\tau^z X\tau^z$. The blue shade represents the $X$-dephasing acting on links with probability $r$.}
\label{Fig_lat}
\end{figure}
\noindent{\underline{Subgroup Ising SSSB in the 2D cluster state:}}
By using a similar prescription to that of the 1D case, we show that a 2D SPT cluster state \cite{Else2012,Yoshida} transforms to a $Z_2$-Ising SSSB state on the vertex lattice. 
The setup of the system is shown in Fig.~\ref{Fig_lat} (b). 
We consider the $L_x\times L_y$ square vertex lattice with periodic boundary conditions and put on link variables between vertices.
Then, the total number of qubits is $3L_xL_y$. 
The lattice is called Lieb lattice. 
Then, Pauli operators residing on vertex sites and links are 
$\tau^\alpha_{j}$ ($\alpha=x,y,z$) and $X_{j+\vec{e}_{\beta}}$, $Y_{j+\vec{e}_{\beta}}$, $Z_{j+\vec{e}_{\beta}}$ ($\beta=x,y$), respectively, where $j$ denotes site and $j+\vec{e}_{\beta}$ a link from the vertex $j$ in $\vec{e}_{\beta}$ direction.

Here, the initial state is set to be the 2D pure cluster state identified with two types of stabilizer, i.e., star stabilizer and link stabilizer, which are given as,  $\rho_{\rm 2DCS}\longleftrightarrow S_{\rm ini}=\{\tau^x_{j}Z_{j+\vec{e}_{x}}Z_{j-\vec{e}_{x}}Z_{j+\vec{e}_{y}}Z_{j-\vec{e}_{y}} | j_{x(y)}=0,1,\cdots, L_{x(y)} \}+\{ \tau^z_{j}X_{j+\vec{e}_x}\tau^z_{j+\hat{x}}, \tau^z_{j}X_{j+\vec{e}_y}\tau^z_{j+\hat{y}} | j_{x(y)}=0,1,\cdots, L_{x(y)}-1 \}$. 
The schematic image of the stabilizers is shown in Fig.~\ref{Fig_lat} (b).
It is known that the 2D cluster state is protected by $Z_2$ 0-form parity symmetry $G^{[0]}_1=\prod_{j}\tau^x_{j}$ and a subgroup one-form symmetry $G^{[1]}_2=\prod_{\gamma \in \mbox{\small loop on link}}X_{\gamma}$, that is $Z^{[0]}_{2}\times Z^{[1]}_2$ symmetry.    
Then, we apply a single-round link $X$-dephasing channel to $\rho_{\rm 2DCS}$, which has the $Z^{[0]}_{2}\times Z^{[1]}_2$ strong symmetries:
\begin{eqnarray}
&&\mathcal{E}^{X}_{\mbox{{\small all link}}}[\rho]= \prod_{j} \mathcal{E}^{X}_{j+\vec{e}_x}\circ\mathcal{E}^{X}_{j+\vec{e}_y}[\rho].\nonumber
\label{2D_entire_dephasing_channel_ZZ}
\end{eqnarray}
By applying the above channel in the stabilizer formalism, the final mixed state is given by
$\mathcal{S}^{ini}\xlongrightarrow{\mathcal{E}^{X}_{\mbox{\small all link}}}
\{G^{[0]}_1\}+\{ \tau^z_{j}X_{j+\vec{e}_x}\tau^z_{j+\hat{x}}, \tau^z_{j}X_{j+\vec{e}_y}\tau^z_{j+\hat{y}} | j_{x(y)}=0,1,\cdots, L_{x(y)}-1 \}
\Longrightarrow \rho^f_{\rm 2DCS}$.
Note that the decohered final state $\rho^f_{\rm 2DCS}$ is stabilized by the symmetry generators $G^{[0]}_1$. 
This subgroup stabilizer generator defined on the vertex sites gives a finite value of R\'{e}nyi-2 correlator $C^{\rm II}_{\tau^z_{i} \tau^z_{j}}(\rho^f_{\rm 2DCS})=1$, while $C^{\rm II}_{\tau^z_{i} \tau^z_{j}}(\rho_{\rm 2DCS})=0$, where $\tau^z_{i}$ is a charged operator of $G^{[0]}_1$. 
Also, $C^{\rm I}_{\tau^z_{i} \tau^z_{j}}(\rho^{(f)}_{\rm 2DCS})=0$. 
Thus, the state $\rho^{f}_{\rm 2DCS}$ is a subgroup SSSB of the zero-form symmetry $G^{[0]}_1$.

\section{Circuit numerical demonstration}
In the previous section, we explained the SSSB and R\'{e}nyi-2 correlation, and gave concrete examples of the SSSB states created by means of the specific single-layer dephasing, which is compose of strong-symmetric projective measurements without monitoring outcomes. 

In this section, we shall consider the issue if there exists a phase transition concerning the SSSB in circuit dynamics. 
The study of such dynamics can give some insight into real experiments. Decoherence in our protocol is nothing but simple quantum noise in the circuit. 
That is, we are interested in how such an effect changes the initial stabilized state into some SSSB state if such a noise can be stochastically controlled.  

To investigate the above issue, we employ stochastic DoCs with a controllable probability. 
By making use of the quantum trajectory picture that is applicable to mixed states \cite{Gullans2020}, we carry out large-scale numerical simulations by using the efficient stabilizer algorithm that is applicable to mixed stabilizer states \cite{Gottesman1997,Aaronson2004}. 
We demonstrate how a noise-induced SSSB transition is realized at the level of the quantum trajectory ensemble of mixed state. 
Furthermore, its criticality will be investigated in detail. 

This study is motivated by the previous works on SSSB phase transitions, which are observed for decohered states emerging under weak measurement (weak decoherence) applied uniformly to the whole system with `probability’ $p$. 
For the 2D systems, an SSSB phase transition was predicted at a finite $p$ by using analytical methods and knowledge of well-known statistical models. 
However, there exists clear difference between the above methods and ours, which will be addressed later after explaining our protocol.

\subsection{Stochastic dephasing-only stabilizer circuit}
We explain the setting of the stochastic stabilizer circuit in detail, where the SSSB can take place in the level of quantum trajectory ensemble. 
We focus on a 2D system and a DoC, and we put local dephasing corresponding to a projective measurement without monitoring (recording) outcomes. 
Such local decoherence using measurement operators corresponding to the Pauli group can be numerically implemented in the algorithm \cite{Weinstein2022}, as explained in Appendix B. 

We consider applying a sequence of local dephasing covering all sites 
in 2D system with a probability $r$ at each site. 
Then, we record locations of local dephasing applied to the system but no other information of the channel. 
Then, a single trajectory of the state labeled by $\{s\}$ is described as
\begin{eqnarray}
\rho^s_{D}=\mathcal{E}^{LD}_{i_0}\circ 
\mathcal{E}^{LD}_{i_1}\circ \cdots \circ \mathcal{E}^{LD}_{i_{N_D}}[\rho_0], 
\end{eqnarray}
where $\rho^s_{D}$ is a finial mixed state, $\rho_0$ is an initial state, $\mathcal{E}^{LD}_{i_k}$ is a local projective measurement at a position $i_k$ without monitoring outcome, and $N_D$ is number of $\mathcal{E}^{LD}_{i_k}$ performed with the probability 
$r$ ($N_D \sim r \times [\mbox{ total number of site}]$)~\cite{Med_come}. 
A similar setup was used in Ref.~\cite{Gullans2020}. 

We produce many samples of the trajectory density matrix, that is, we prepare realizations of the random dephasing $\mathcal{E}^{LD}_{i_k}$ patterns as many as possible. 
By using these samples of the density matrix, we calculate the ensemble average of R\'enyi-2 correlator given by 
\begin{eqnarray}
\overline{C^{\rm II}_{O_i O^\dagger_j}}&\equiv&
\mathbb{E}[C^{{\rm II},s}_{O_i O^\dagger_j}],\label{barCII}\\
C^{{\rm II},s}_{O_i O^\dagger_j}&\equiv& \frac{\Tr[O_i O^\dagger_j\rho^s_D O_j O^\dagger_i \rho^s_D]}{\Tr[(\rho^s_D)^2]},
\label{R2_ens}
\end{eqnarray}
where $\mathbb{E}[\cdot]$ denotes averaging over the samples of the trajectory density matrix. 
The practical calculation methods of Eq.~(\ref{R2_ens}) in the stabilizer formalism is explained in Appendix C. 
Note that each sample of the density matrix satisfies the strong symmetry condition, thus, we can employ the notion and framework of the SSSB. 
For pure states, $\overline{C^{\rm II}_{Z_i Z^\dagger_j}}$ (called Edward-Anderson spin glass order) has been used in the context of the measurement-induced phase transition \cite{Sang2021}.   

We remark that the observable $\overline{C^{\rm II}_{O_i O^\dagger_j}}$ in Eq.~(\ref{barCII}) is different from the quantity defined by $$\displaystyle{C^{\rm II}_{Z_i Z^\dagger_j}\equiv \frac{\Tr[Z_iZ_j\rho_D Z_jZ_i \rho_D]}{\Tr[(\rho_D)^2]}},$$
where $\rho_D$ is the density matrix obtained through the decoherence channel without recording locations and outcomes of performed local dephasings. 
Explicitly for the Ising dephasing case in Sec. III A, the general local decoherence operator is given as 
$\rho \to (1-p)\rho+pZ_iZ_{i+1}\rho  Z_iZ_{i+1}$,
where $p$ is the dephasing parameter and $\mathcal{E}^{Z_iZ_{i+1}}$ in 
Eq.~(\ref{entire_dephasing_channel_ZZ}) corresponds to $p=1/2$. 
\textit{The uniform product of the above local operations at all links generates $\rho\to \rho_D$. }
The parameter $r$ in our protocol seems to be closely related to $p$ (such as $p=r/2$), but there exists a sharp difference between them.
A simple example of this difference is shown in Appendix D.
From the above perspective, we would like to comment on the difference between our study and the analysis of the previous works, in which the density matrix with dephasing is analytically treated by mapping it to a pure state: (I) In our protocol, we record locations of operated local dephasing, (II) We focus on physical quantities calculated for each sample (circuit trajectory) and average them over generated samples.

In the rest of the paper, we shall calculate the R\'{e}nyi-2 susceptibility and its trajectory average defined as 
\be
\chi^{\rm II,s}&=&{1 \over V}\sum_{i,j}C^{\rm II,s}_{O_iO^\dagger_j}, 
\label{chaiIIs}  \\
\overline{\chi^{\rm II}}  &\equiv& \mathbb{E}[\chi^{\rm II,s}], \label{barchiII}
\ee
where $V$ is  the size of the system, $V=L_xL_y$, and $\{s\}$ is the label of samples again.
Further quantity considered is its variance;
$$\sigma\equiv {\rm var}[\chi^{\rm II,s}].
$$
This quantity is useful to observe the SSSB phase transition.
Please see the following subsections.
\begin{figure}[t]
\begin{center} 
\vspace{0.5cm}
\includegraphics[width=8.8cm]{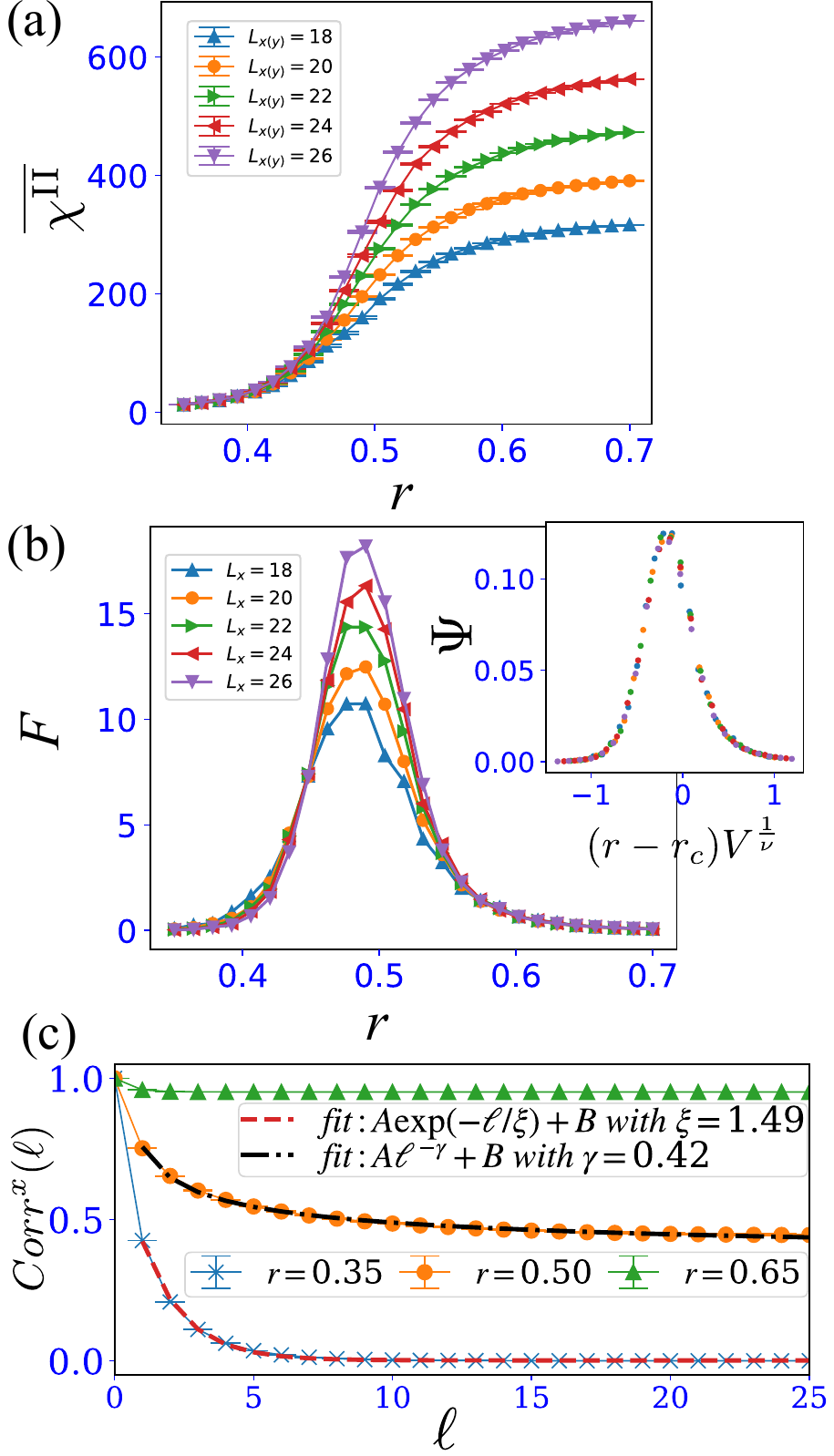}  
\end{center} 
\caption{(a) Behavior of $\overline{\chi^{\rm II}}$ for various values of $r$. 
(b) Behavior of $\sigma /V\equiv F$ as varying $r$. 
The inset panel is a scaling function, where we extract the critical transition point and its criticality, $r_c=0.51$, $\nu = 3.01\pm 0.10$ 
and $\zeta = 2.32\pm 0.11$. 
For each point data of (a) and (b), we used 2000 samples (trajectories). 
(c) Spatial profile of R\'enyi-2 correlator with various r. We used the following fitting functions:
$A\exp(-\ell/\xi)+B$ and $A\ell^{-\gamma}+B$, where $A, B, \xi$, and $\gamma$ are fitting parameters. 
As for panel (c), we took 2000 sample disorder average, and the system size is fixed as $L_{x(y)}=50$.}
\label{Fig_0}
\end{figure}

\subsection{2D stochastic Ising SSSB}
Now we show the first numerical study for the stochastic DoC. 
We consider 2D $L_x \times L_y$ square lattice with periodic boundary conditions, and qubits reside on lattice sites. 
The total number of sites is $V=L_x L_y$. 
The target symmetry is $Z_2$ parity $P_{Z_2}=\prod_{\rm all: (i_x,i_y)}X_{(i_x,i_y)}$.
We prepare the initial $+X$ product state, $\rho_{\rm ini}=|{\bf X}\rangle \langle {\bf X}|$, 
where $|{\bf X}\rangle=|+\rangle^{\otimes V}$ and $|+\rangle$ is $+1$ eigenstate of Pauli $X$.
Here, we consider the dephasing on a link, which is already introduced in the previous section and given by $\mathcal{E}^{Z_{(i_x,i_y)}Z_{(i_x+1,i_y)}}$ or $\mathcal{E}^{Z_{(i_x,i_y)}Z_{(i_x,i_y+1)}}$. 
We apply the $x$ and $y$-directed dephasing to each link with probability $r$ and obtain the single trajectory sample of the final mixed state, $\rho^s_{\rm fin}$. 
All prescriptions in that process are strong symmetric for the parity $P_{Z_2}$.

By setting $O_{(i_x,i_y)}=Z_{(i_x,i_y)}$, we numerically calculate $\overline{\chi^{\rm II}}$ and $\sigma /V\equiv F$ by varying the probability $r$. 
The obtained results of $\overline{\chi^{\rm II}}$ are shown in Fig.\ref{Fig_0} (a). 
We see that for $r<0.4$, data obtained for all system sizes have quite small values, whereas for $r\gtrsim 0.5$ all data start to increase and they exhibit different behavior depending on the system size. 
In order to investigate the critical behavior, we display $F$ in Fig.\ref{Fig_0} (b). 
We find the sharp peak in $F$ as a function of $r$, and as the system gets larger, the peak becomes larger. 
The value of $r$ at the peak for each system is close to $r\sim 0.5$. 
These results indicate the existence of an SSSB phase transition. 
 
To locate the transition point and its criticality, we perform the finite-size scaling analysis to $F(r,V)$ using pyfssa numerical package~\cite{Pyfssa1,Pyfssa2}. 
The scaling ansatz is set as $F(r,V)=(V)^{\frac{\zeta}{\nu}}\Psi((r-r_c)(V)^{\frac{1}{\nu}})$, where $\Psi$ is a scaling function, $\zeta$ and $\nu$ are critical exponents and $r_c$ is a critical transition probability in the thermodynamics limit.

We observe the clear data collapse as shown in the inset of Fig.~\ref{Fig_0}(b). Here, the critical transition probability is estimated by $r_c=0.51$ and the exponents, $\nu = 3.01\pm 0.10$ and 
$\zeta = 2.32\pm 0.11$. 
The above scaling analysis confirms the SSSB phase transition induced solely by the stochastic dephasing in our protocol.
It should be remarked that the obtained critical value $r_c$ is in good agreement with the threshold of the 2D bond percolation ($= 0.50$) \cite{Aharony2003}. 
This observation is discussed in Sec.~VI rather in detail. 
On the other hand, the exponent $\nu$  is related to the correlation length of the Edward-Anderson (EA) type for `spins' $\{Z_{i_x,i_j}\}$, as the R\'{e}nyi-2 correlator reduces to that of the EA for the pure state.
Although the exponent $\nu$ does not have a definite interpretation in 
the percolation perspective, we can discuss its possible relation to the 2D percolation. 
The discussion is given by Appendix E.

It is interesting to observe the spatial correlation of the R\'enyi-2 correlator given by
\begin{eqnarray}
{\rm Corr}^x(\ell)\equiv\mathbb{E}[C^{{\rm II},s}_{Z_{(i_x,i_y)}Z_{(i_x+\ell,i_y)}}],
\label{Corrx}
\end{eqnarray}
where $0<\ell\leq L_x/2$.
The numerical results for $L_x=L_y=50$ is shown in Fig.\ref{Fig_0} (c). 
Depending on the value of $r$, different behavior of the correlations is clearly observed. 
At the critical point, the power law decay is the best fit for ${\rm Corr}^x(\ell)$, and below the critical transition point, ${\rm Corr}^x(\ell)$ exhibits an exponential decay \cite{fitting_justification}. 
Above the critical transition point, the correlation remains for 
$\ell \to$ large.
These results obviously support the observation that the behavior of $F$ indicates the existence of the SSSB
transition emerging on varying $r$.

\subsection{2D stochastic subgroup Ising SSSB}

\begin{figure}[t]
\begin{center} 
\vspace{0.5cm}
\includegraphics[width=8.9cm]{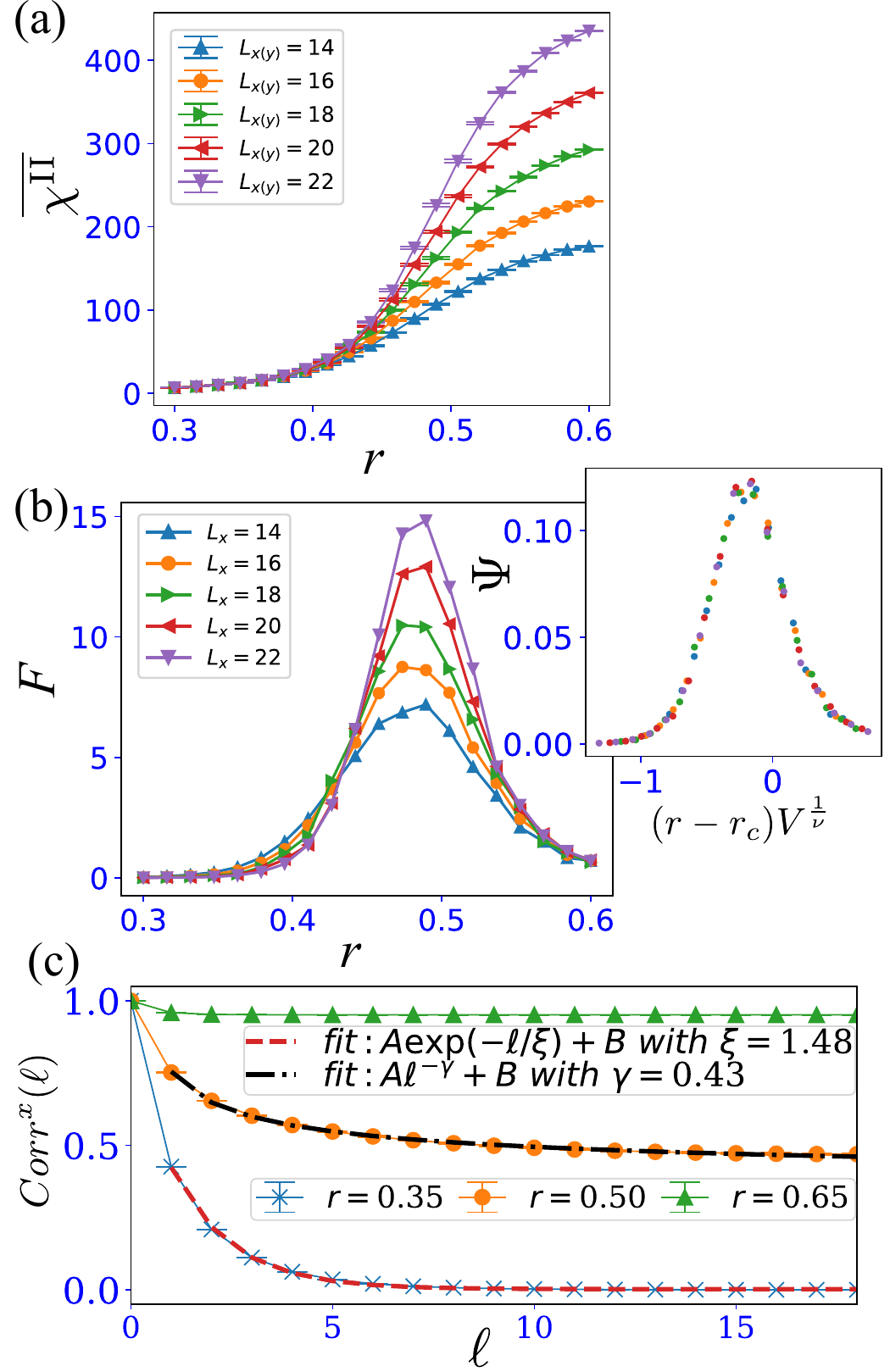}  
\end{center} 
\caption{
(a) Behavior of $\overline{\chi^{\rm II}}$  for various values of $r$.
(b) Behavior of $\sigma /V\equiv F$ as varying $r$. 
The inset panel is a scaling function, where we extract the critical transition point and its criticality, $r_c=0.51 \pm 0.02$, $\nu = 2.96\pm 0.23$ 
and $\zeta = 2.30\pm 0.35$. 
For each point data of (a) and (b), we used 2000 samples (trajectories).
(c) Spatial profile of R\'enyi-2 correlator with various r. We used the following fittig functions:
$A\exp(-\ell/\xi)+B$ and $A\ell^{-\gamma}+B$, where $A, B$, $\xi$, and $\gamma$ are fitting parameters. 
As for panel (c), we took 1000 sample disorder average, and the system size is fixed as $L_{x(y)}=36$.}
\label{Fig_1}
\end{figure}
We show the second concrete numerical study for the stochastic DoC;
the 2D cluster state on the Lieb lattice shown in Fig.~\ref{Fig_lat} (b). 
The total number of vertices is $V=L_x L_y$.
The setup is the same as the previous one.

Here, the initial density matrix is set to the 2D pure cluster state, $\rho_{\rm 2DCS}\longleftrightarrow S_{\rm ini}=
\{\tau^x_{j}Z_{j+\vec{e}_{x}}Z_{j-\vec{e}_{x}}Z_{j+\vec{e}_{y}}Z_{j-\vec{e}_{y}}| j_{x(y)}=0,1,\cdots, L_{x(y)}-1 \}+\{ \tau^z_{j}X_{j+\vec{e}_x}\tau^z_{j+\hat{x}}, \tau^z_{j}X_{j+\vec{e}_y}\tau^z_{j+\hat{y}}| j_{x(y)}=0,1,\cdots, L_{x(y)}-1 \}$. 
Then, we apply a single round of a link $X$-dephasing channel with probability $r$ to the initial state.  
The local link one is $\mathcal{E}^{X}_{j+\vec{e}_x}\circ\mathcal{E}^{X}_{j+\vec{e}_y}[\rho]$, which respects the $Z^{[0]}_{2}\times Z^{[1]}_2$ symmetry. 
Therefore, all manipulations in the DoC preserve the $Z^{[0]}_{2}\times Z^{[1]}_2$ symmetry. 

By setting $O_j=\tau^z_j$ with $j \in \mbox{[vertex]}$, 
we again numerically calculate $\overline{\chi^{\rm II}}$ and $F$ by varying the probability $r$. 
The result of $\overline{\chi^{\rm II}}$ is shown in Fig.\ref{Fig_1} (a). 
We see that for $r<0.4$, all data for various system sizes exhibit quite small values,
whereas, for $r\gtrsim 0.5$, all data start to increase to finite values and their behavior depends on the system size. 
Then, $F$ exhibits a sharp peak as displayed in Fig.\ref{Fig_1} (b). 
The location of the peak for each system is close to $r\sim 0.5$, indicating the existence of a phase transition there.

To locate the critical point and clarify the criticality, we again perform the finite-size scaling analysis to $F(r,V)$ with the same scaling ansatz to the previous case. 
We observe the clear data collapse as shown in Fig.~\ref{Fig_1}(b). 
Here, the critical transition probability is estimated by $r_c=0.51 \pm 0.02$ and the exponents, $\nu = 2.96\pm 0.23$ and $\zeta = 2.30\pm 0.35$.  
Thus, the above scaling analysis confirms the SSSB phase transition induced solely by the stochastic dephasing. 
The critical point is close to that of 2D bond percolation ($= 0.50$) \cite{Aharony2003}. 
On the other hand, although the exponent $\nu$ does not have a clear interpretation in the percolation perspective, we can discuss its possible relation to the 2D percolation as  the case in Sec.~IV.B.
For details, please see  Appendix E.

Finally, we show the spatial behavior of the R\'enyi-2 correlator that is given by
\begin{eqnarray}
{\rm Corr}^x(\ell)\equiv\mathbb{E}[C^{{\rm II},s}_{Z_{(i_x,i_y)}Z_{(i_x+\ell,i_y)}}],
\end{eqnarray}
where $0<\ell\leq L_x/2$.
The numerical results for $L_x=L_y=36$ are shown in Fig.\ref{Fig_1} (c). 
Almost similar behavior to the previous case is observed. 
In the vicinity of the critical point, the power law decay appears, while below the critical transition point, the correlation exhibits exponential decay \cite{fitting_justification}. 
\begin{figure*}[t]
\begin{center} 
\vspace{0.5cm}
\includegraphics[width=17.5cm]{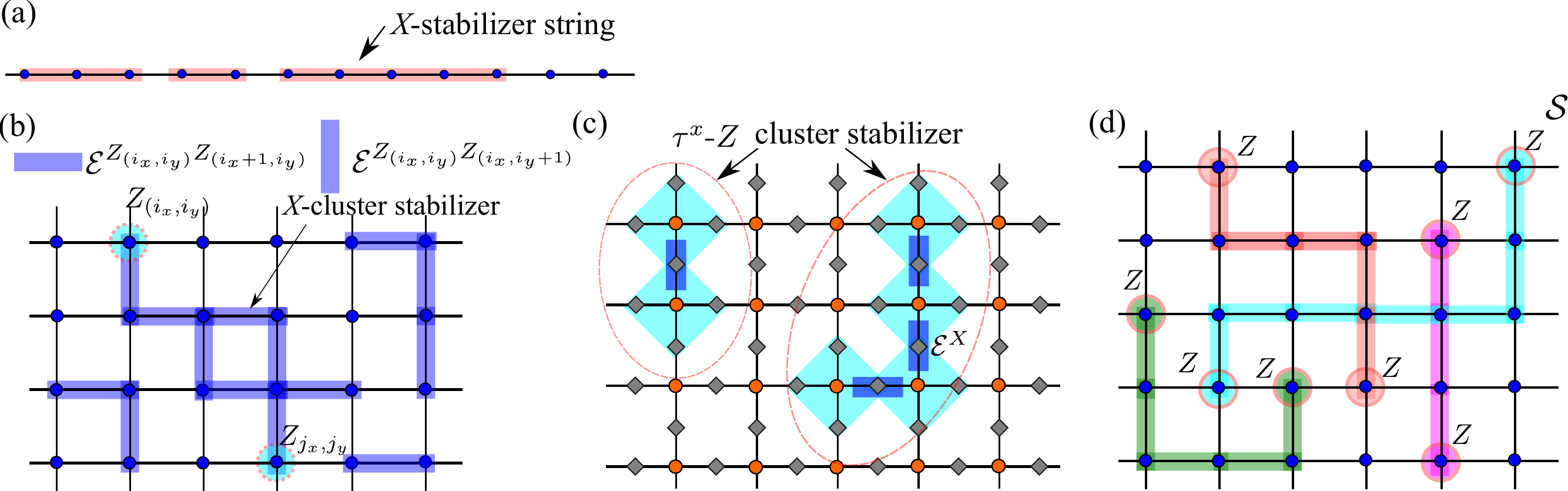}  
\end{center} 
\caption{
(a) Schematic image of $X$-stabilizer string generated by the dephasing where the 1D bond percolation picture is useful. If the $X$-product string covers the two separate charged operators in the R\'{e}nyi-2 correlation, the R\'{e}nyi-2 correlation has a finite value.
(b) Schematic image of $X$-cluster stabilizer generator generated by the dephasing where the 2D bond percolation picture is useful. The blue shade located on links represent $ZZ$-link dephasing. The light blue circles represent a pair of charged operators for the R\'{e}nyi-2 correlator.
When a $X$- cluster stabilizer covers a pair of charged operators in the R\'{e}nyi-2 correlator, a finite value of the R\'{e}nyi-2 correlation emerges. Otherwise, it vanishes. 
(c) Schematic image of $\tau^x$-$Z$-cluster stabilizer generator generated by the dephasing where the 2D bond percolation picture is useful. 
The light blue diamonds represent original $\tau^x$-$Z$ stabilizers in Fig.~\ref{Fig_lat}(b), 
the connection of which becomes $\tau^x$-$Z$-cluster stabilizer generator generated by the dephasing. 
The dephasing is represented by the blue shade on links. When a cluster stabilizer covers a pair of separate charged operators located on center site of blue diamonds, the corresponding R\'{e}nyi-2 correlator has a finite value.
(d) Schematic image of a string configuration $\{\mathcal{S}\}$ in a $X$-cluster stabilizer generator (its boundary is not shown.).
}
\label{Fig_parc}
\end{figure*}

\section{Comment on 1D stochastic SSSB state}
We comment if some stochastic DoC in a 1D system exhibits a phase transition or not. 
We conclude that genuine SSSB states similar to those discussed for the 2D systems can exist but no phase transition takes place. 
With a similar setup to the 2D cases, physical quantities characterizing the SSSB such as the R\'enyi-2 correlator and its susceptibility exhibit only smooth change as varying $r$. 
In fact, the SSSB takes place only for the specific case with $r=1$, that is, the miximal decoherence channel as discussed in \cite{Lessa2024,Sala2024}, and the origin of this phenomenon will be explained in Sec.~VI by using percolation picture.

\section{Spontaneous strong-to-weak symmetry breaking and percolation picture}

In Sec.~IV, we displayed numerical calculations for the SSSB transitions in the 2D models. 
In this section, we shall examine the obtained results from a geometrical point of view, that is, percolation picture in the square lattice. 
Let us focus on the 2D Ising dephasing for concreteness. 
 As we demonstrated in Sec.~III A, the local dephasing operation of $(Z_{i_x,i_y}Z_{i_x+1,i_y})$ merges the original stabilizers $\{X_{i_x,i_y},X_{i_x+1,i_y}\} $ to the combined one $(X_{i_x,i_y}X_{i_x+1,i_y})$, and so on. 
 The density matrix after dephasing is given as 
 $\rho \propto \prod_{\ell} (1+g_{\ell})$
 with the resultant stabilizer generators $\{g_\ell\}$. 
 The R\'{e}nyi-2 correlator is calculated as 
$$
{\rm Tr} \biggl[Z_{i_x,i_y}Z_{j_x,j_y}\prod_\ell  (1+g_\ell) Z_{i_x,i_j}Z_{j_x,j_y} \prod_{\ell'} (1+g_{\ell'})\biggr], 
$$
and therefore, R\'{e}nyi-2 correlator has a non-vanishing value when $Z_{i_x,i_y}Z_{j_x,j_y}$ is commutative with all the stabilizers. 
This observation reveals important notion like `X-cluster stabilizer’, 
and when sites $i$ and $j$ both belong to support of one of $X$-cluster stabilizers,  
${\rm Tr} [Z_{i_x,i_y}Z_{j_x,j_y} \prod_\ell (1+g_\ell) Z_{i_x,i_y}Z_{j_x,j_y} \prod_{\ell'} (1+g_{\ell'})] >0$,
and otherwise zero. 
In other words, sites $(i_x,i_y)$ and $(j_x,j_y)$ must be connected by a line of measured bonds for a finite value of R\'{e}nyi-2 correlator.
Similar consideration can be applied to the 1D system, and it also easily derives the conclusion that in 1D $ZZ$ dephaseing system, no SSSB states appear except $r=1$ as a finite value of R\'{e}nyi-2 correlation for $|i-j| \to \infty$ requires the infinitely long string (cluster in 1D) residing the whole system ~\cite{2D_parc}. 
A schematic image is shown in Fig.~\ref{Fig_parc} (a).

The above observation is depicted in Fig.~\ref{Fig_parc} (b), which shows that the 2D $ZZ$ dephasing is similar to the process of the link percolation, whereas 2D cluster is close to the bond percolation as shown in Fig.~\ref{Fig_parc} (c). 
For the 2D percolation, the threshold is known as $\sim 1/2$ \cite{Aharony2003},
which is in a good agreement with the numerically obtained values in the previous section.

This geometrical and pictorial interpretation of the decohered mixed states in the stabilizer formalism can be corroborated by the doubled Hilbert space prescription~\cite{J_Y_Lee2022}, in which emergent decohered mixed states are mapped to pure states and their string representation is available. Let us denote the initial pure state as $|\Omega_0\rangle$, and then the corresponding density matrix is given by 
$\rho_0 = |\Omega_0\rangle\langle \Omega_0|$.
The doubled Hilbert space is composed of two copies of the original Hilbert space, and  the above density matrix is mapped to a pure state in the doubled Hilbert space as 
$\rho_0 \to ||\rho_0\rangle\rangle \equiv |\Omega_0\rangle_u|\Omega_0\rangle_l$, where the subscript $(u,l)$ refers to the up and low Hilbert space, respectively.
For the 2D ZZ dephasing case, $|\Omega_0\rangle=|+\rangle^{\otimes V}$, let us consider general form of decoherence channel  
such as 
$$
\mathcal{E}:\rho \to (1-p)\rho + p Z_vZ_{v'}\rho Z_vZ_{v'},
$$
where $v$ and $v'$ denote site of the square lattice and $p$ is the dephasing parameter, $0<p\le  1/2$.
In the doubled Hilbert space formalism, the Choi operator corresponding to the uniform application 
of the above decoherence to the whole system is given as
\be
\hat{\mathcal{E}} = \prod_{(v,v')\in NN} (1-2p)^{1/2}
e^{\tau Z_{v, u}Z_{v',u} Z_{v, l}Z_{v',l}},
\label{Choi1}
\ee
where $\tanh \tau =p/(1-p)$ and $NN$ stands for nearest neighbor sites.
The above operator in Eq.~(\ref{Choi1}) is invariant independently under $\prod_uX$ and $\prod_lX$ reflecting the strong symmetry of the original
operation on the mixed state.
The limit $p\to 1/2$  means $\tau\to \infty$, and therefore, the dephasing prescription in the our protocol corresponds to the low temperature limit and maximal dephasing.

In our stochastic dephasing prescription, we first choose randomly links on which dephasing operation works.
This is a kind of quenching disorder, and an emergent configuration of operated links gives a sample $s$,
which \textit{is composed of a set of cluster stabilizers} discussed in the above.
[Link configuration of a sample determines a set of cluster stabilizers in that sample.]
For each emergent sample $s$, Choi operator $\hat{\mathcal{E}}_s$ generates the following strong symmetric state in the doubled Hilbert space in the string representation,
\be
||\rho^{D}_s\rangle\rangle \equiv
\hat{\mathcal{E}}_s||\rho_0\rangle\rangle \propto
\sum_{\mathcal{S}\in s} (\tanh \tau)^{{|\mathcal{S}|}}
|\partial \mathcal{S}\rangle_u \otimes |\partial \mathcal{S}\rangle_l,
\label{Choi2}
\ee
where $\mathcal{S}\in s$ denotes \textit{all possible string configurations} included in sample 
$s$, and  
$|\partial \mathcal{S}\rangle_u =\prod_{v \in \partial\mathcal{S}}Z_{v,u}|\Omega_0\rangle_u$ and similarly for $|\partial \mathcal{S}\rangle_l$.
A typical example of string ensemble $\{\mathcal{S}\}$ is shown in Fig.~\ref{Fig_parc}~(d),
and $\{\mathcal{S}\}$ obviously corresponds to a \textit{ X-cluster stabilizer generator} in our prescription.
The R\'{e}nyi-2 correlator is obtained in the doubled Hilbert space by calculating 
the expectation value $\langle\langle Z_{v_1,u}Z_{v_1,l}Z_{v_2,u}Z_{v_2,l}\rangle\rangle_s$
with respect to $||\rho^{D}_s\rangle\rangle$ in Eq.~(\ref{Choi2}) by taking the limit $\tanh \tau \to 1$, i.e., no dampings. 
Then, R\'{e}nyi-2 correlation emerges between any pair of sites $(v_1,v_2)$ if there exist strings  $\{\mathcal{S}\}$  connecting  $(v_1,v_2)$.
This condition is equivalent to that $(v_1,v_2)$ belong to the support of one of X-cluster stabilizer generators
in sample $s$.

The above consideration clarifies that the present stabilizer protocol for the stochastic dephasing is a 
kind of quench dynamics, and instead of assigning a finite dephasing parameter $p$, sparse links are created 
with probability $(1-r)$.
This point of view might explain the fact that the critical exponent obtained by the numerical calculation
in the previous section is close to that of the 2D percolation model. 
Anyway, further study is need to get clear understanding of the critical behavior of the present circuit dynamics.

\section{Conclusion and discussion}
In this paper, we studied dephasing/decoherence effects on quantum many-body systems. 
The SSSB phenomena was concisely explained and some concrete examples of the SSSB state were given. 
The SSSB phenomena broadly emerge in various physical systems, giving rich classification schema for states of matter in open quantum systems. 

We further studied the SSSB by using the stochastic DoC, which has possibility to clarify the location of the phase transition of SSSB by observing the averaged R\'enyi-2 correlator and quantum trajectory of mixed state samples simulated by numerical methods. 
We focused on two systems, (I) 2D square lattice system with local ZZ-link dephasing and (II) 2D Lieb lattice system where 2D cluster state is chosen as an initial state and local $X$-link dephasing is performed.   
We found the clear phase transitions in the above 2D systems and clarified their critical properties by using finite-size scaling etc.  
We discussed the physical picture of the phase transitions by using the similarity between the present quantum systems and percolation.
We hope that this approach will shed light into other systems that exhibit an SSSB phase transition.

Finally, we shall give a perspective about a possible SSSB of a higher-form symmetry by considering toric code \cite{Kitaev} as an example, Hamiltonian of which is given by
\be
H&=&-\sum_e\prod_{e\in v}X_e-\sum_p\prod_{e\in p}Z_e \nonumber \\
&=&-\sum_v A_v-\sum_p B_p,
\ee
where $Z_e$ and $X_e$ reside on link (edge) of the square lattice with vertex (site) $v$, and the periodic boundary conditions ($T^2$ torus) are imposed.
There are two kinds of one-form symmetries generated by 
$\prod_{e\in \gamma} Z_e$ and $\prod_{e\in \tilde{\gamma}} X_e$ \cite{McGreevy2022,Verresen2022}, where $\gamma$ is an arbitrary loop residing links and $\tilde{\gamma}$ is an arbitrary loop crossing links.
The ground state is four-fold degenerate and satisfy $A_v=B_p=1$.
In particular, we choose a state out of the ground state multiplet such as
$\prod_{e\in {\cal C}_i}Z_e$=1, where ${\cal C}_i \; (i=1,2)$ are two non-contractible loops on the torus.
We consider the decoherence under the channel, the local element of which is given as; 
$\mathcal{E}_e:\rho \to (1-p)\rho+pX_e\rho X_e$.

We can apply a Kramers-Wannier duality to the ground state, which is defined by introducing the 2D dual lattice; site of dual lattice 
$\tilde{v}$  corresponds to plaquette $p$ of the original lattice, link of the original lattice $e$ is expressed by a pair of sites of the dual lattice $(\tilde{v},\tilde{v}')$. 
By the duality, operators are related as follows; 
$\prod_{e\in p}Z_e \Longleftrightarrow X_{\tilde{v}}$ and $X_e \Longleftrightarrow Z_{\tilde{v}}Z_{\tilde{v}'}$ \cite{J_Y_Lee2023}.
Then, we expect that the above ground state of toric code is nothing but $|+\rangle^{\otimes V}$ on the dual lattice, and also the decoherent channel is that of Ising ZZ system.
Therefore, the results obtained by studying the decohered 2D Ising channel provide the SSSB structure of the decohered toric code ground state.
The most important observation is that the R\'{e}nyi-2 correlator defined by Eq.~(\ref{Corrx})  
is transformed to a R\'{e}nyi-2 correlator of the ‘t Hooft string, which is defined as $H_{\tilde{\Gamma}}=\prod_{e\in\tilde{\Gamma}}X_e$
with an open string $\tilde{\Gamma}$ crossing links, an order parameter of the magnetic one-form symmetry.
[The distance $\ell$ in Eq.~(\ref{Corrx})  corresponds to the length of $\tilde{\Gamma}$.]
Transition to the SSSB state in the dephasing 2D Ising system means that the magnetic one-form-symmetry order emerges;
$\langle H_{\tilde{\Gamma}} \rho^D H_{\tilde{\Gamma}}\rho^D\rangle \neq 0$ and $\langle  \rho^D H_{\tilde{\Gamma}}\rangle = 0$.  
In fact, a percolation picture can be also applicable to this system.
The detail will be reported in future work.

We expect that there are further various examples and extensions of the SSSB state. 
One of the interesting directions of future studies is the application of the notion of the SSSB to some higher-form symmetries \cite{Gaiotto2015,McGreevy2022}, where toric code with or without open boundary conditions is the simplest example as we discussed in the above very briefly, and there are relevant works on 
this subject in Refs.~\cite{Verresen2022,KOI2024}.   
 
\section*{Acknowledgements}
This work is supported by JSPS KAKENHI: JP23K13026(Y.K.) and JP23KJ0360(T.O.). 
We acknowledge the anonymous referee for much helpful suggestions on the interpretation of the critical exponents shown in Sec.IV B and C and in Appendix E. 

\appendix

\section*{Appendix A: Standard transformation in a set of stabilizer generator}
Quantum state is identified by a set of stabilizer generators. 
This set is not unique \cite{Nielsen_Chuang}. 
The set of the stabilizer generators is denoted by $\{g_0,\cdots, g_{N-1}\}$, where $N$-independent generators are considered. 
As explained in Ref.~\cite{Nielsen_Chuang}, there is a standard transformation between the stabilizers. 
The set of stabilizer generators are multiplying $g_i$ by $g_j$ ($i\neq j$) to obtain a new stabilizer generator $g_{i}\to g_{i}g_j\equiv g'_{i}$. 
In this transformation, the stabilizer group obtained from stabilizer generators is invariant. 
This rule can include the sign of the stabilizer generators even though we can ignore the sign factor throughout this work. 
We can construct a tractable set of stabilizer generators to identify the corresponding many-body states. 
This prescription works similarly for the stabilizer generators with the outcome factors $g_i\to \beta_jg_{i}$ with $\beta_j=\pm 1$. 
In the standard transformation, we can change the form of the stabilizer generators by multiplying $\beta_ig_i$ with $\beta_j g_j$ ($i\neq j$) to obtain a transformed stabilizer generator as  $\beta_i g_{i}\to \beta_i\beta_jg_{i}g_j\equiv \beta_i\beta_j g'_{i}$. 

\section*{Appendix B: Update rule of dephasing in stabilizer formalism}
Throughout this work, we consider local dephasing corresponding to measurements with the measuring operator $\hat{m}_{i}$ without monitoring (recording) the outcomes, regarded as a local strong decoherence channel. 
Here, we assume that $\hat{m}_{i}$ is an element of Pauli group with $+1$ factor, where the outcome denoted by $\beta_i$ is $\beta_i=\pm 1$. 
The channel of this local dephasing is given by 
\begin{eqnarray}
\mathcal{E}^{\hat{m}}_i[\rho]=\sum_{\beta_i=\pm}P^{m_i}_{\beta_i}\rho P^{m_i\dagger}_{\beta_i}=\frac{1}{2}\rho+\frac{1}{2}\hat{m}_i\rho \hat{m}_{i},
\label{local_dephasing_channel}
\end{eqnarray}
where $P^{m_i}_{\beta_i}$ is a projection operator of $\hat{m}_i$ with outcome $\beta_i$, $P^{m_i}_{\beta_i}=\frac{1+\beta_i \hat{m}_i}{2}$.
The entire channel is represented as
\begin{eqnarray}
\mathcal{E}^{\hat{m}}[\rho]= \prod^{L-1}_{i=0} \mathcal{E}^{\hat{m}}_i[\rho].
\label{entire_dephasing_channel}
\end{eqnarray}

Let us see how the dephasing acts to a mixed state practically.  
We consider that the density matrix is represented by stabilizer generators $\{g_{\ell}\}$,  
\begin{eqnarray}
\rho=\frac{1}{2^{L-k}}\prod^{k-1}_{\ell=0}\frac{1+g_{\ell}}{2}.
\label{dens_stab}
\end{eqnarray}
According to Refs.\cite{Weinstein2022}, the introduction of the local dephasing channel $\mathcal{E}^{\hat{m}}_i$ in the stabilizer formalism is efficiently implemented in the stabilizer algorithm. 
The prescription is the following;
When one applies $\mathcal{E}^{\hat{m}}_i$ to $\rho$, then the density matrix represented by the stabilizer generator is
\begin{eqnarray}
\mathcal{E}^{\hat{m}}_i[\rho]&=&\sum_{\beta_i=\pm}P^{m_i}_{\beta_i}\rho P^{m_i\dagger}_{\beta_i}\nonumber\\
&=&\sum_{\beta_i=\pm}P^{m_i}_{\beta_i} \biggl[\frac{1}{2^{L-k}}\prod^{k-1}_{\ell=0}\frac{1+\tilde{g}_{\ell}}{2}\biggr]P^{m_i\dagger}_{\beta_i}\nonumber\\
&=&\sum_{\beta_i=\pm}P^{m_i}_{\beta_i} \biggl[\frac{1+\tilde{g}_{0}}{2}\biggr]P^{m_i\dagger}_{\beta_i} \biggl[\frac{1}{2^{L-k}}\prod^{k-1}_{\ell=1}\frac{1+\tilde{g}_{0}}{2}\biggr]\nonumber\\
&=&\frac{1}{2^{L-k+1}}\prod^{k-1}_{\ell=1}\frac{1+\tilde{g}_{\ell}}{2},
\end{eqnarray}
where on the second line we have performed a standard transformation (See Appendix A) to change the representation of the set of stabilizer generators $\{g_\ell\}$ 
into the one denoted by $\{\tilde{g}_\ell\}$, in which at most one stabilizer generator labeled 
by $\tilde{g}_{0}$ is anticommute with $\hat{m}_i$. 

Thus, application of the local dephasing $\mathcal{E}^{\hat{m}}_i$ eliminates a few stabilizer generators from the set of stabilizer generators, leading to the enhancement of the mixing of the state.

\section*{Appendix C: Calculation method of R\'enyi-2 correlator}
The R\'enyi-2 correlator can be calculated in the stabilizer formalism. 
We consider a charged operator product $O_i$ for a symmetry with a symmetry charge. 
We assume that the charged operator belongs to the Pauli group including identity, and the density matrix is given by Eq.~(\ref{dens_stab}).   
Since each stabilizer $g_{\ell}$ commutes or anti-commutes with the charged operator, the R\'enyi-2 correlator for a density matrix $\rho$ is given as 
\begin{eqnarray}
&&C^{\rm II}_{\rm O_iO_j}[\rho]=\frac{\Tr[O_iO_j\rho O_jO_i\rho]}{\Tr[\rho^2]},
\end{eqnarray}
and the numerator and denominator are calculated as 
\begin{eqnarray}
&&\Tr[O_i O_j\rho O_j O_i\rho]=\frac{1}{2^{L}}\biggl[\prod^{k-1}_{\ell=0}(1+\alpha^\ell)\biggl],\\
&&\Tr[\rho^2]=\frac{1}{2^{L-k}},
\end{eqnarray}
with the factor $\alpha^\ell=\pm$ for $[O_iO_j,g_\ell]_{\pm}=0$ (where $[\cdot]_{\pm}$ is commutative or anti-commutative bracket) and we have used $O_iO_j(1+g_{\ell})O_jO_i=(1+\alpha^{\ell}g_{\ell})$. 
Thus, we only extract the (anti)-commutativity between $O_iO_j$ and each stabilizer generator to obtain the R\'enyi-2 correlator.

\section*{Appendix D: Channel dynamics with local stochastic decoherence without monitoring locations}
We consider the 2D Ising dephasing channel, in which the DoC does not record locations of  operated dephasing. 
The density matrix on the circuit is given by  
\begin{eqnarray}
&&\rho^{D}_{\rm wr}\equiv E^{r}_{\rm tot}[\rho_{\rm ini}] = \prod_{(i_x,i_y)}E^{\hat{x}}_{(i_x,i_y)}\circ E^{\hat{y}}_{(i_x,i_y)}
[\rho_{\rm ini}],\nonumber\\
&&E^{\hat{x}(\hat{y})}_{(i_x,i_y)}[\rho_{\rm ini}]=(1-r)\rho_{\rm ini}+r\mathcal{E}^{Z_{(i_x,i_y)}Z_{(i_x+1,i_y (i_y+1))}}[\rho_{\rm ini}].\nonumber
\end{eqnarray}
Note that there is a difference in the physical quantities $\hat{A}$ calculated with the two density matrices $\rho^s_D$ and 
$\rho^{D}_{\rm wr}$.

Some physical quantities given by the linear form of the density matrix $\rho^D_{\rm wr}$ are the same. 
For example, 
$$
\Tr[\hat{A}\rho^D_{\rm wr}\hat{A}^\dagger]=\mathbb{E}[A^s]
$$
with $A^s= \Tr[\hat{A}\rho^s_D \hat{A}^\dagger]$.
That is, the two different density-matrix formulations give the same results since the observable is given by the linear form of the density matrix.
However, physical quantities denoted by $\hat{B}$ obtained by a nonlinear form of density matrix are {\it not identical} such as 
$$
\Tr[\hat{B}\rho^D_{\rm wr}\hat{B}^\dagger\rho^D_{\rm wr}]\neq \mathbb{E}[B^s]
$$
with 
$B^s= \Tr[\hat{B}\rho^s_D \hat{B}^\dagger \rho^s_D]$.
Even for the simplest case $\hat{B}=\hat{I}$ (which is a purity), the two quantities obtained from the two different density matrices become different.

\section*{Appendix E: Possible relation for critical exponents}
In Sec.~IV B, we carried out the scaling analysis by employing the ansatz, 
$F(r,V)=(V)^{\frac{\zeta}{\nu}}\Psi((r-r_c)(V)^{\frac{1}{\nu}})$. 
Whereas, there is more standard ansatz such as, $F(r,V) \to F'(r,L)\equiv L^{\frac{\zeta '}{\nu '}}\Psi((r-r_c)L^{\frac{1}{\nu '}})$, 
where $L$ is a certain side-length of the system, and the exponent $\nu'$ is nothing but the genuine critical exponent of the correlation length. 
For the numerical calculations in Sec.~IV B, $F'(r,L)$ is readily obtained from $F(r,V)$ by putting $L=L_x(=L_y)$. 
Then, we obtain $\nu'= \frac{1}{2}\nu \sim 1.5$, which is fairly close to the value of the 2D percolation $\nu_{2DP}=4/3$. 
On the other hand, we also obtain $\frac{\zeta '}{\nu '}\equiv \kappa \sim 1.16$ to find that there is a certain amount of difference from the value of corresponding quantity of 2D percolation, $\kappa_{2DP}=43/24$.  
This result requires detailed investigation how the R\'{e}nyi-2 correlator and some correlators in the 2D percolation theory (square lattice) are related. 
This issue is an interesting future work. The same discussion is applicable to the exponents obtained in 
Sec.~IV.C. 
There, by using the scaling ansatz $F(r,V) \to F'(r,L)\equiv L^{\frac{\zeta '}{\nu '}}\Psi((r-r_c)L^{\frac{1}{\nu '}})$, we also see that $\nu '\sim 1.48$, close to $\nu_{2DP}=4/3$.



\begin{thebibliography}{99}
\bibitem{stab_GHZ}
In the stabilizer formalism, the state has only $Z_2$ parity operator $\prod_j X_j$.
\bibitem{stab_1DIsing}
Even in absence of a single local dephasing in link channel $\mathcal{E}^{Z_i Z_{i+1}}$, the final state $\rho_{\rm fin}$ does not change.
\bibitem{inf_T}
On the other hand, if we apply $\mathcal{E}^{Z}_{\rm all}$, which is a weak symmetric channel for the $Z_2\times Z_2$ symmetry \cite{Groot},
the final state is just an infinite temperature state $\mathcal{E}^{Z}_{all}[\rho_{CS}]=\frac{1}{2^{L}}I$, and the information of the initial SPT state is totally washed out.
\bibitem{Med_come}
One significant advantage of this circuit is its experimental-friendly setting. Each trajectory is identified only by the locations where local dephasing is applied. Therefore, we can realize the same trajectory by applying dephasing on the same locations.
\bibitem{fitting_justification}
By checking a residual error we have numerically evaluated the validity of each fitting function. 
Based on the residual error, we found that the exponential fit is more suitable for $r< 0.5$, while the power-law fit is more appropriate for $r\sim0.5$.
\bibitem{2D_parc}
In 2D system, we can take many possible $X$-cluster stabilizer generators generated by some sequence of the dephasing covering the operators $Z_{i_x,i_y}Z_{j_x,j_y}$. Thus, the R\'{e}nyi-2 correlation tends to have a finite value while in 1D case, there is only one-directed X-string stabilizer generator.
\end{thebibliography}
\end{document}